\documentclass[11pt, a4paper]{article}
\pdfoutput=1
\usepackage[authoryear]{natbib}
\usepackage{amsmath,amssymb,amsfonts,amsthm,bbm}
\usepackage{epic,eepic,epsfig,longtable}
\usepackage{multirow,verbatim}
\usepackage{array}
\usepackage{pdflscape}
\usepackage{wasysym}
\usepackage{algorithm}
\usepackage[colorlinks,citecolor=blue,linkcolor=blue]{hyperref}
\usepackage{epsfig}
\usepackage{setspace}
\usepackage{mathrsfs}
\usepackage[figuresleft]{rotating}
\usepackage{amsbsy}
\usepackage{eqnarray}
\usepackage{booktabs}
\usepackage{epstopdf}
\usepackage{color}

\makeatletter
\def\singlespace{\def\baselinestretch{1}\@normalsize}

\makeatletter
\def\singlespace{\def\baselinestretch{1}\@normalsize}

\numberwithin{equation}{section}

\newcommand{\bfm}[1]{\ensuremath{\mathbf{#1}}}

\newcommand{\be}{\bfm e}     
\newcommand{\bff}{\bfm f}  \newcommand{\bF}{\bfm F}

   \newcommand{\bI}{\bfm I}

   \newcommand{\bM}{\bfm M}

   \newcommand{\bS}{\bfm S}

   \newcommand{\bW}{\bfm W}  
   \newcommand{\bX}{\bfm X}  
\newcommand{\by}{\bfm y}   \newcommand{\bY}{\bfm Y}  
   \newcommand{\bZ}{\bfm Z}  

 \newcommand{\cA}{{\cal  A}}

 \newcommand{\cM}{{\cal  M}}
 \newcommand{\cN}{{\cal  N}}
 \newcommand{\cO}{{\cal  O}}
 \newcommand{\cP}{{\cal  P}}

 \newcommand{\cS}{{\cal  S}}

 \newcommand{\cX}{{\cal  X}}

\newcommand{\bfsym}[1]{\ensuremath{\boldsymbol{#1}}}
 
 \newcommand{\bbeta}{\bfsym \beta}
              \newcommand{\bGamma}{\bfsym \Gamma}

 \newcommand{\bmu}{\bfsym {\mu}}                 
 \newcommand{\bnu}{\bfsym {\nu}}

              \newcommand{\bSigma}{\bfsym \Sigma}
         \newcommand{\bLambda}{\bfsym {\Lambda}}
           \newcommand{\bOmega}{\bfsym {\Omega}}

\newcommand{\C}{\mbox{const.}\quad}

               \newcommand{\hbSigma}{\hat{\bfsym \Sigma}}
             
             \newcommand{\hbLambda}{\hat{\bfsym \Lambda}}

\DeclareMathOperator{\Cov}{Cov}

\DeclareMathOperator{\E}{E}

\DeclareMathOperator{\tr}{tr}

\def \tr     {\mathrm{tr}}
\def \det    {\mathrm{det}}
\def \logdet {\mathrm{logdet}}

\def \onf    {\mathrm{1,off} }
\def \E      {\mathrm{E}}
\def \cO     {\mathcal{O}}

\def\MSE{\mbox{MSE}}

\def\Cov{\mbox{Cov}}
\def\C{\mbox{const.}\quad}

\allowdisplaybreaks
\usepackage{verbatim}
\renewcommand{\baselinestretch}{1.27}
\newtheorem{theorem}{Theorem}
\newtheorem{remark}{Remark}
\newtheorem{proposition}{Proposition}
\newcounter{CondCounter}

\begin{document}

\title{\textbf{ On the penalized maximum likelihood estimation of high-dimensional approximate factor model} \thanks{ The work was supported by a project of Shandong Province Higher Educational Science and Technology Program (Grant No. J17KA160) and the National Natural Science Foundation of China (Grant Nos. 11761020, 11671059) }}
\author{
  Shaoxin Wang$^a$\thanks{Corresponding author. Email: shxwang@qfnu.edu.cn or shwangmy@163.com (S.X. Wang), hy@cqu.edu.cn(H. Yang)}, Hu Yang$^b$, Chaoli Yao$^b$ \vspace{0.2in}\\
\small $^a$School of Statistics, Qufu Normal University, Qufu 273165, P.R. China\\
\small $^b$College of Mathematics and Statistics, Chongqing University, Chongqing 401331, P.R. China}

\date{Last modified on \today}

\maketitle
\begin{abstract}
  In this paper, we mainly focus on the penalized maximum likelihood estimation (MLE) of the high-dimensional approximate factor model. Since the current estimation procedure can not guarantee the positive definiteness of the error covariance matrix, by reformulating the estimation of error covariance matrix and based on the lagrangian duality, we propose an accelerated proximal gradient (APG) algorithm to give a positive definite estimate of the error covariance matrix. Combined the APG algorithm with EM method, a new estimation procedure is proposed to estimate the high-dimensional approximate factor model. The new method not only gives positive definite estimate of error covariance matrix but also improves the efficiency of estimation for the high-dimensional approximate factor model. Although the proposed algorithm can not guarantee a global unique solution, it enjoys a desirable non-increasing property.  The efficiency of the new algorithm on estimation and forecasting is also investigated via simulation and real data analysis.
\end{abstract}


\section{Introduction}
\label{intro}
The factor model finds its popularity in psychometrics, economics and many other research fields for the utility in summarizing information in large dataset. Many researches have been done to investigate different topics relating to the factor model, such as its estimation and statistical inference theory \citep{Mula09, Bai03}, the dynamic factor model and its generalization \citep{Gewe78,Forn00}, and another active topic related to the factor model is covariance matrix estimation, whose recent development can be found in the survey articles by \citet{BaiS11}, \citet{FanL16} and the reference therein.

Let $\by=(y_1,\cdots,y_p)^T$ be a $p$-dimensional observable random vector. Then, the factor model \citep{Mula09} can be stated as follows
\begin{equation}\label{eq.FctMdl}
    \by=\bmu+\bLambda\bff+\be,
\end{equation}
where $\bff=(f_1,\cdots,f_r)^T$ is a vector of common factors of the $p$ random variables in $\by$, and $\bLambda$ is the $p\times r$ common factor loading matrix. $\bff$ and $\be=(e_1,\cdots,e_p)^T$ are unobservable random vectors and uncorrelated with each other. $e_i$ is the idiosyncratic error of $y_i$ correspondingly. Thus, the covariance matrix of $\by$, $\Cov(\by)=\bSigma_{\by}$, can be decomposed as
\begin{equation}\label{decomSgmy}
    \bSigma_{\by}=\bLambda \Cov(\bff)\bLambda^T+\bSigma_{\be}.
\end{equation}
When the covariance matrix of $\be$, $\Cov(\be)=\bSigma_{\be}$, is a diagonal matrix, model \eqref{eq.FctMdl} is called the strict factor model. In studying the arbitrage in market, \citet{Ross76} showed that when the strict factor model is satisfied, the mean $\bmu$ of assets is approximately linear function of factor loadings. \citet{Cham83} relaxed the diagonal assumption and showed that as long as the eigenvalue of $\bSigma_{\be}$ is bounded the conclusion sill holds. This leads to the approximate factor model that allows the dependence among the error terms. In this paper, we mainly focus on the approximate factor model. More researches on the strict factor model are referred to \citet{FanFL08}, \citet{Mula09}, \citet{Hiro15}  and so on.

Many estimation procedures of the approximate factor model have been developed. By restricting the error covariance matrix to be diagonal, the consistency properties and limiting distribution of the maximum likelihood (ML) estimators of approximate factor model have been established by \cite{BaiS11} and \cite{DozG2012}. \citet{Choi12} proposed the generalized principal component (PC) analysis method.  Recently,  in high-dimensional setting \cite{Bai16} proposed an penalized
maximum likelihood (PML) method to estimate the factor loading matrix and the error covariance matrix simultaneously. In addition to the diagonal elements, the PML method also gives a sparse estimate of the off-diagonal elements of $\bSigma_{\be}$. But, we note that the PML method can not guarantee the estimate of $\bSigma_{\be}$ to be positive (semi)definite and even the positive definiteness of $\bSigma_{\by}$ can not be ensured. And this stimulates our work. It should be pointed out that the result of the PML method is closely related to the high-dimensional covariance matrix estimation problem in which the covariance matrix has the structure \eqref{decomSgmy}.  With different procedures, \citet{FanLM11} and \cite{FanL13} first gave sparse estimates of the $\bSigma_{\be}$, and then the covariance matrix $\bSigma_{\by}$. More researches on the estimation of high-dimensional covariance matrix are referred to \citep{Bick08C,CaiL11,Roth08,Roth09,LiuW14,CuiL16,XueM12,BaiS11,FanL16}. To achieve a sparse estimate of $\bSigma_{\be}$, penalty functions or thresholding techniques should be used. Examples include LASSO and it variants \citep{Tibs96,Zou06}, the folded concave penalization \citep{FanL01,Zhang10}, and more details on the high-dimensional data analysis should be referred to \cite{BuhlV11}.

In this paper, we only consider the penalized MLE of the high-dimensional approximate factor model, which gives simultaneous estimates of the factor loading matrix $\bLambda$ and the error covariance matrix $\bSigma_{\be}$. A detailed comparison of the ML based and PC based estimation procedures for the approximate factor model has been given in \citep{Bai16}, and the superiorities of the ML based method are also given there. As we have said that the PML method used in \citep{Bai16} can not guarantee a positive definite estimate of the covariance matrix. To overcome this drawback forms the main contents of our paper. By reconsidering the PML method, we propose a new estimation equation and the corresponding algorithm. Our method not only gives simultaneous estimates of $\bLambda$ and $\bSigma_{\be}$ but also guarantees the positive definiteness of $\hbSigma_{\be}$. Similar to \citep{Bai16}, we also check the influence of our new estimators on the weighted least squares (WLS) estimate of the factor $\bff$. Of course there are also many other factor extraction methods \citep[Ch.~7]{Mula09}, but for comparison and its simplicity we only consider the WLS method. As a byproduct, we also show that our method can also be used to estimate the covariance matrix $\bSigma_{\by}$, which was not addressed in \citep{Bai16}.

The rest of the paper is organized as follows. In Section \ref{sect.2} we revisit the PML method and propose our estimation function. The algorithm for solving our estimation equation is proposed in the Section \ref{sect.3}.  All the simulation and a real data analysis are presented in Section \ref{sect.4}. The concluding discussion and the proof of main results are given in Section \ref{sect.5} and Appendix, respectively.

\section{Methodology}
\label{sect.2}
We assume that $\bff$ and $\be$ in model \eqref{eq.FctMdl} are nondegenerate multivariate normal random vectors with $\E(\bff)=\mathbf{0}$ and $\E(\be)=\mathbf{0}$.
Let $\by_1,\by_2,\cdots, \by_n$ be the random observations from \eqref{eq.FctMdl}. If we further assume that $\by_1,\by_2,\cdots, \by_n$ are mutually independent, then the negative log-likelihood function is given by
\begin{align*}
  L(\bSigma_{\by}) &=\frac{n}{2} \logdet(\bSigma_{\by})+\frac{n}{2}\tr\left(\bSigma_{\by}^{-1} \bS_{\by}\right)+\frac{n}{2}p\log(2\pi),
\end{align*}
where we use $\logdet(A)$ to denote $\log(\det(A))$ for simplicity, and $\bS_{\by}=\frac{1}{n}\sum^{n}_{j=1}(\by_j-\bar{\by})(\by_j-\bar{\by})^T$ with $\bar{\by}=\frac{1}{n}\sum^{n}_{j=1}\by_j$. We take \eqref{decomSgmy} into the above equation and get
\begin{align*}
L(\bLambda,\bSigma_{\be})& =\logdet\left(\bLambda\Cov(\bff)\bLambda^T+\bSigma_{\be}\right)+\tr\left((\bLambda\Cov(\bff) \bLambda^T+\bSigma_{\be})^{-1}\bS_{\by}\right)+\C
\end{align*}
Note that the MLE of $\bLambda$ is largely depending on the estimation of $\bSigma_{\be}$. However, in high-dimensional setting the estimation of $\bSigma_{\be}$ is difficult due to the fact that estimating too many parameters $\cO(p^2)$ with a relative small sample size $n$. Therefore, additional structure assumption on $\bSigma_{\be}$ is usually needed. One typical and wildly used assumption is that $\bSigma_{\be}$ is conditional sparse, which means many off-diagonal elements are zero or nearly zero \citep{Bai16}. To get a parsimonious estimate of $\bSigma_{\be}$, penalty function will be employed to shrink the small off-diagonal elements of $\bSigma_{\be}$ to be zero. Discarding the mutual independence assumption among the observations, \citet{Bai16} proposed the following penalized quasi-maximum likelihood (PML) estimation of the approximate factor model.

\subsection{The PML method}
In the PML estimation of the approximate factor model, some identification conditions are needed to estimate $\bLambda$. With the following restriction \citep{Lawl71, Bai16} that
\begin{equation}
\label{Id.C}
 \Cov(\bff)=\bI_r,\quad \bLambda^T\bSigma_{\be}^{-1}\bLambda\textrm{ is diagonal,}
\end{equation}
and the elements of $\bLambda^T\bSigma_{\be}^{-1}\bLambda$ are distinct and arranged in descending order, \citet{Bai16} proposed the following PML estimation equation to estimate $\bLambda$ and $\bSigma_{\be}$
\begin{align}\label{PMLEbaiL}
  (\hbLambda,\hbSigma_{\be})& = \mathop\mathrm{argmin}_{\bLambda,\bSigma_{\be}} \log\left(\left|\mathrm{det}(\bLambda \bLambda^T+\bSigma_{\be})\right|\right)+\tr\left((\bLambda \bLambda^T+\bSigma_{\be})^{-1}\bS_{\by}\right)+\mathrm{P}_{\lambda}(\bSigma_{\be}),
\end{align}
where $|\cdot|$ is to take the absolute value of the determinant of $\bLambda \bLambda^T+\bSigma_{\be}$. $\mathrm{P}_{\lambda}(\bSigma_{\be})$ is a penalty function and given by
\begin{eqnarray*}
  \mathrm{P}_{\lambda}(\bSigma_{\be}) &=& \lambda\|\bW\circ \bSigma_{\be}\|_{\onf},
\end{eqnarray*}
where $\lambda$ is the tuning parameter, $\bW$ is a matrix of weights, the symbol $\circ$ denotes the Hadamard product, and $\|\cdot\|_{\onf}$ sums up the absolute value of the off-diagonal elements. {Let $w_{ij}$ denote the element of $\bW$. When $w_{ij}=1$, the well known LASSO penalty \citep{Tibs96} follows. Let $\hbSigma_{\be,ij}$ be a preliminary consistent estimate of the element in $\bSigma_{\be}$. If we set $w_{ij}=\hbSigma_{\be,ij}^{-1}$, then $\mathrm{P}_{\lambda}(\bSigma_{\be})$ corresponds to the adaptive LASSO penalty \citep{Zou06}. Furthermore if we set
\begin{align*}
w_{ij}=\left[I_{(|\hat{\rho}_{\be,ij}|\leq \lambda)}+\frac{(c-|\hat{\rho}_{\be,ij}|/\lambda)_{+}}{c-1}I_{(|\hat{\rho}_{\be,ij}|> \lambda)}\right]\times(\hbSigma_{\be,ii}\times \hbSigma_{\be,jj})^{-\frac{1}{2}}
\end{align*}
with $\hat{\rho}_{\be,ij}=\hbSigma_{\be,ij}/(\hbSigma_{\be,ii}\times \hbSigma_{\be,jj})^{1/2}$, then the smoothly clipped absolute deviation (SCAD) penalty \citep{FanL01} follows.}

We note that when $\bLambda \bLambda^T+\bSigma_{\be}$ is nonsingular, taking the absolute value of its determinant makes \eqref{PMLEbaiL} well defined. But this can not guarantee the estimate of $\bSigma_{\by}$, $\hbSigma_{\by}=\hbLambda\hbLambda^T+\hbSigma_{\be}$, to be positive definite. Since $\bSigma_{\be}$ is the covariance matrix of the error term, it should be positive definite or at least positive semidefinite. If we get a positive definite estimate of $\bSigma_{\be}$, then it is sufficient for $\hbSigma_{\by}$ to be positive definite. Motivated by this we propose the following method to estimate the approximate factor model.

\subsection{The proposed method}

Under the same identification condition \eqref{Id.C}, we adopt the following constrained PML estimation equation to estimate the factor loading matrix and the error covariance matrix of model \eqref{eq.FctMdl}
\begin{align}
\label{eq.minPnlhood}
  (\hbLambda,\hbSigma_{\be})=&\mathop\mathrm{argmin}_{\bLambda,\bSigma_{\be},\mathrm{ and }\bSigma_{\be}\succ 0}L(\bLambda,\bSigma_{\be}),
\end{align}
where
\begin{align*}
L(\bLambda,\bSigma_{\be})=\logdet\left(\bLambda \bLambda^T+\bSigma_{\be}\right)+\tr\left((\bLambda \bLambda^T+\bSigma_{\be})^{-1}\bS_{\by}\right)+\mathrm{P}_{\lambda}(\bSigma_{\be}),
\end{align*}
and $\bSigma_{\be}\succ 0$ means $\bSigma_{\be}$ is positive definite. Compared with \eqref{PMLEbaiL}, we do not need to take the absolute value of the determinant of $\bLambda \bLambda^T+\bSigma_{\be}$. Meanwhile, our method not only gives simultaneous estimates of $\bLambda$ and $\bSigma_{\be}$ but also guarantees the estimate of $\bSigma_{\be}$  to be positive definite. If we assume the observations are independent and drop the constraint $\bSigma_{\be}\succ 0$, \eqref{eq.minPnlhood} is just the penalized MLE equation of the approximate factor model. After obtaining $\hbLambda$ and $\hbSigma_{\be}$, with \eqref{decomSgmy} the estimate of $\bSigma_{\by}$ can be easily obtained by
$$\hbSigma_{\by}=\hbLambda\hbLambda^T+\hbSigma_{\be},$$
which can be treated as an easy byproduct of our estimation procedure and was not discussed in \citep{Bai16}.

\section{Algorithm}
\label{sect.3}
The EM algorithm \citep{McKr07} has been widely used in the literature to find the MLE of the factor models \citep{Rubin82,Hiro16}. In this section, we propose an EM based iterative procedure to give the estimators of $\bLambda$ and $\bSigma_{\be}$ in the approximate factor model \eqref{eq.FctMdl}.

Let $\bF=[\bff_1,\cdots,\bff_n]$ be treated as the missing data, $\bY=[\by_1,\cdots,\by_n]$ be the observed one. Then, given $\bY$ and the $k$-th estimates of $\bLambda$ and $\bSigma_{\be}$, denoted by $\bLambda^{(k)}$ and $\bSigma_{\be}^{(k)}$, the conditional expectation of the penalized log-likelihood function of complete data is
\begin{align}
\label{Estep}
   \E(L_c(\bLambda,\bSigma_{\be})|\bY) =&-\frac{n}{2}\logdet(\bSigma_{\be})-\frac{n}{2}\tr(\bSigma_{\be}^{-1}\bS_{\by})+ n\tr(\bLambda^T\bSigma_{\be}^{-1}\bS_{\by}\bGamma^{(k)})-\frac{n}{2}\mathrm{P}_{\lambda}(\bSigma_{\be}) \nonumber\\
   & -\frac{n}{2}\tr\left((\bLambda^T\bSigma_{\be}^{-1}\bLambda+\bI_r)({\bGamma^{(k)}}^T\bS_{\by}\bGamma^{(k)} +\bOmega^{(k)})\right)+\C,
\end{align}
where $\bGamma^{(k)}={\bSigma_{\by}^{(k)}}^{-1}\bLambda^{(k)}$ with $\bSigma_{\by}^{(k)}=\bLambda^{(k)}{\bLambda^{(k)}}^T + {\bSigma_{\be}}^{(k)}$, and $\bOmega^{(k)}=\bI_r-{\bLambda^{(k)}}^T\bGamma^{(k)}$. The derivation of \eqref{Estep} is given in Appendix part \ref{sect.A}. The $k+1$-th estimates of $\bLambda$ and $\bSigma_{\be}$ are given by
\begin{align}
\label{Mstep}
\left(\bLambda^{(k+1)}, {\bSigma_{\be}}^{(k+1)}\right)=&\mathop\mathrm{argmin}_{\bLambda,\bSigma_{\be},\mathrm{ and }\bSigma_{\be}\succ 0} \logdet(\bSigma_{\be})+\tr\left(\bSigma_{\be}^{-1}\bLambda \bOmega^{(k)}\bLambda^T\right)\nonumber\\
&+\tr\left(\bSigma_{\be}^{-1}\left(\bI_p-\bLambda {\bGamma^{(k)}}^T\right)\bS_{\by}\left(\bI_p-\bLambda {\bGamma^{(k)}}^T\right)^T\right)+\mathrm{P}_{\lambda}(\bSigma_{\be}),
\end{align}
which is solved with the following iterative procedure.
\begin{description}
  \item[Step 1.] Treating $\bSigma_{\be}$ as a constant matrix, we take the derivative of \eqref{Mstep} with respect to $\bLambda$ and update $\bLambda$ as follows
\begin{equation}
\label{UpdateLmd}
\bLambda^{(k+1)}=\bS_{\by}\bGamma^{(k)}\left(\bOmega^{(k)}+{\bGamma^{(k)}}^T\bS_{\by}\bGamma^{(k)}\right)^{-1}.
\end{equation}
  \item[Step 2.] Substituting $\bLambda^{(k+1)}$ into equation \eqref{Mstep}, and $\bSigma_{\be}$ is updated by
\begin{align}
\label{UpdateSgme}
{\bSigma_{\be}}^{(k+1)}&=\mathop\mathrm{argmin}_{\bSigma_{\be}\succ 0} \logdet(\bSigma_{\be})+\tr\left(\bSigma_{\be}^{-1}\bM\right)+\mathrm{P}_{\lambda}(\bSigma_{\be}),
\end{align}
where
\begin{align}\label{MatM}
    \bM=&\left(\bI_p-\bLambda^{(k+1)} {\bGamma^{(k)}}^T\right)\bS_{\by}\left(\bI_p-\bLambda^{(k+1)} {\bGamma^{(k)}}^T\right)^T+\bLambda^{(k+1)} \bOmega^{(k)}{\bLambda^{(k+1)}}^T.
\end{align}
Here, we need to claim that the $\bLambda$s in $\bGamma$ and $\bOmega$ are also updated. Since $\bSigma_{\be}^{(k)}$ is still unchanged, we here just write $\bGamma^{(k)}$ and $\bOmega^{(k)}$ for simplicity.
\end{description}
We note that equation \eqref{UpdateSgme} has the same form as the penalized MLE of covariance matrix discussed by \citet{Bien11}, except for the matrix $\bM$. And the nonconvexity makes it challenging to solve \eqref{UpdateSgme}. From equation \eqref{MatM}, it can be easily checked that when $\bS_{\by}$ is positive definite, $\bM$ is also positive definite. According to Proposition 1 \citep{Bien11}, it can be easily deduced that when $\bM$ is positive definite, the constraint $\bSigma_{\be}\succ 0$ can be tightened to $\bSigma_{\be}\succ \delta \bI_p$ for some $\delta>0$. Thus, when $\bS_{\by}$ is positive definite, \eqref{UpdateSgme} is equivalent to \begin{align*}
{\bSigma_{\be}}^{(k+1)}&=\mathop\mathrm{argmin}_{\bSigma_{\be}\succ \delta\bI_{p}} \logdet(\bSigma_{\be})+\tr\left(\bSigma_{\be}^{-1}\bM\right)+\mathrm{P}_{\lambda}(\bSigma_{\be}).
\end{align*}
 Following the MM procedure \citep{Lang16} and the proximal gradient method used in \citep{Bien11}, we give an alternative to Step 2 as follows.
\begin{description}
  \item[Step $\mathbf{2^{\prime}}$.] Choose a suitable depth of projection $t$, and update $\bSigma_{\be}$ as follows
\begin{equation}
\label{approxObj}
 \bSigma_{\be}^{(k+1)}=\mathop\mathrm{argmin}_{\bSigma_{\be}\succeq \delta \bI_p}\frac{1}{2t}\left\|\bSigma_{\be}-\cM_n\right\|_{F}^{2}+\mathrm{P}_{\lambda}(\bSigma_{\be}),
\end{equation}
where $\cM_n=\bSigma_{\be}^{(k)}-t\left({\bSigma_{\be}^{(k)}}^{-1}- \left({\bSigma_{\be}^{(k)}}^{-1}\bM{\bSigma_{\be}^{(k)}}^{-1}\right)\right)$.
\end{description}
The details on the derivation of \eqref{approxObj} can be found in \citep{Bien11}. We need to point out that when $p>n$, $\bS_{\by}$ (or $\bM$) is only positive semidefinite and the constraint $\bSigma_{\be}\succ \delta\bI_{p}$ will not hold. {If we drop the constraint, \eqref{approxObj} reduces to the following equation
\begin{equation}
\label{approxObjnoC}
 \bSigma_{\be}^{(k+1)}=\mathrm{arg}\min\frac{1}{2t}\left\|\bSigma_{\be}-\cM_n\right\|_{F}^{2}+ \mathrm{P}_{\lambda}(\bSigma_{\be}),
\end{equation}
which was used by \citet{Bai16} to update $\bSigma_{\be}$. We can check that \eqref{approxObjnoC} has a closed-form solution which exactly corresponds to the generalized thresholding covariance matrix estimator \citep{Roth09}. We also note that the $\bSigma_{\be}^{(k+1)}$ given by \eqref{approxObjnoC} may contain negative eigenvalue due to the thresholding procedure.  A similar discussion on the high-dimensional covariance matrix estimation has been given by \citet{XueM12}, which asserted that the generalized thresholding covariance matrix estimator can not guarantee the estimate of covariance matrix to be positive definite.} So in high-dimensional setting we directly give an lower bound $\delta$ of the minimum eigenvalue of $\bSigma_{\be}$, which forms the constraint $\bSigma_{\be}\succeq \delta \bI_p$ in \eqref{approxObj}. It should be noted that $\delta$ is not a tuning parameter  and  we may choose $\delta=10^{-4}$. Similar treatment has been used in high-dimensional covariance matrix estimation to guarantee a positive definite estimate of the covariance matrix \citep{Roth12,XueM12,LiuW14}. \citet{LiuW14} also proposed a data driven procedure to determine the value of $\delta$.

To solve \eqref{approxObj}, many methods, like alternating direction method of multipliers \citep[ADMM,][]{Boyd11}, can be used. However, when the ADMM is taken, a penalty parameter in the augmented lagrange function needs the user choose. The penalty parameter has no influence on the theoretical convergence results \citep{XueM12}, but can give nonnegligible impact on the numerical performance \citep{MaXZ13}. Here, we follow \citet{CuiL16} and solve \eqref{approxObj} by applying the APG algorithm to its lagrangian dual problem.

The lagrange function of \eqref{approxObj} is
\begin{align}\label{Lagran}
L(\bSigma_{\be},\bZ)=&\frac{1}{2t}\left\|\bSigma_{\be}-\cM_n\right\|_{F}^{2}+\mathrm{P}_{\lambda}(\bSigma_{\be})-\tr\left( \bZ^T(\bSigma_{\be}-\delta \bI_p)\right),
\end{align}
where $\bZ$ is the lagrange multiplier. The lagrange dual problem of \eqref{Lagran} is given by
\begin{equation}\label{Lagdp}
    \max_{\bZ\succeq 0}\min_{\bSigma_{\be}}L(\bSigma_{\be},\bZ).
\end{equation}
Let $X(\bZ)=\cM_n+t\bZ$. We define
\begin{align}
\label{g(Z)}
  g(\bZ)\equiv-\min_{\bSigma_{\be}}\left\{L(\bSigma_{\be},\bZ)\right\}
    = &-\inf_{\bSigma_{\be}} \left\{\frac{1}{2t}\left\|\bSigma_{\be}-X(\bZ)\right\|_{F}^{2}+\mathrm{P}_{\lambda}(\bSigma_{\be})\right\}\nonumber\\
  &-\frac{1}{2t}\left\|\delta \bI_p-\cM_n\right\|_{F}^{2} +\frac{1}{2t}\left\|X(\bZ)-\delta \bI_p\right\|_{F}^{2}.
\end{align}
Since $\mathrm{P}_{\lambda}(\bSigma_{\be}) = \lambda\|\bW\circ \bSigma_{\be}\|_\onf$, the first part of \eqref{g(Z)} has a colsed-form solution and its minimizer is given as
\begin{align}
\label{Thresh_Z}
\hbSigma_{\be}&= \cS\left(X(\bZ),t\lambda \bW\right)=\left[\mathrm{sign}(X(\bZ)_{ij})\max\left\{|X(\bZ)_{ij}|-t\lambda \bW_{ij},0\right\}\right],
\end{align}
and $\cS(\cdot,\cdot)$ is the elementwise soft-thresholding operator. Thus for \eqref{g(Z)} we have
\begin{align*}
    g(\bZ)=&-\frac{1}{2t}\left\|\hbSigma_{\be}-X(\bZ)\right\|_{F}^{2}-\mathrm{P}_{\lambda}\left(\hbSigma_{\be}\right)
    -\frac{1}{2t}\left\|\delta \bI_p-\cM_n\right\|_{F}^{2}+\frac{1}{2t}\left\|X(\bZ)-\delta \bI_p\right\|_{F}^{2}.
\end{align*}
By Proposition \ref{Prop1} (see Appendix \ref{MYR}), the function $g(\bZ)$ is continuously differentiable and its gradient is given by
\begin{equation*}
    \nabla g(\bZ)=\cS\left(X(\bZ),t\lambda \bW\right)-\delta \bI_p.
\end{equation*}
Meanwhile, the lagrange dual problem \eqref{Lagdp} can be written as
\begin{equation}\label{Obj_Grad}
    \min_{\bZ} f(\bZ)\equiv g(\bZ)+\delta_{psd}(\bZ),
\end{equation}
where $\delta_{psd}(\cdot)$ is the indictor function of the cone of positive semidefinite matrices. \citet{Beck09} proposed an APG algorithm to deal with \eqref{Obj_Grad}, and proved a complexity result of $\cO(1/k^2)$. Then, from Part 2 of Proposition \ref{Prop1}, $\nabla g(\bZ)$ is globally Lipschitz continuous, and according to the APG algorithm we can solve \eqref{Obj_Grad} by an iterative procedure of minimizing its quadratic approximation
\begin{align}\label{quadratic_Apprx}
    \bY^{(k+1)}&=\mathrm{arg}\min_{\bZ\succeq 0}g(\bZ^{(k)})+\frac{1}{2}\left\|\bZ-\bZ^{(k)}\right\|_F^2+\tr\left(\nabla g(\bZ^{(k)})^T(\bZ-\bZ^{(k)} )\right).
\end{align}
It is well known that the solution to \eqref{quadratic_Apprx} is given by
\begin{equation}\label{}
    \bY^{(k+1)}=\cP_{+}\left(\bZ^{(k)}-\nabla g(\bZ^{(k)})\right),
\end{equation}
where $\cP_{+}(\cdot)$ is the projection operator onto the cone of positive semidefinite matrices. For a symmetric matrix $\bX$, let $\bX=\sum^{p}_{i=1}\lambda_iv_iv_i^T$ be its eigendecomposition, and then its projection onto the cone of positive semidefinite matrices is given by $\cP_{+}(\bX)=\sum^{p}_{i=1}\max\left\{\lambda_i,0\right\}v_iv_i^T$. Now we summarize the above discussion as the following APG algorithm to update $\bSigma_{\be}$ in \textbf{Step $2^\prime$}.
\begin{algorithm*}[htbp]
\caption{\textbf{ Update $\bSigma_{\be}$ by APG algorithm}}\label{APG}
Given $\bSigma_{\be}^{(k)}$, the updated estimate $\bLambda^{(k+1)}$, and $t^{(k)}=1.$ Set $k:=1$. Iterate until convergence:
\begin{description}
  \item[Step $\mathbf{2^\prime} a$] Compute $\bM$ by \eqref{MatM} and $\cM_{n}$ in \eqref{Lagran}.
  \item[Step $\mathbf{2^\prime} b$] Compute $\cS\left(X(\bZ^{(k)}),t\lambda \bW\right)$ by \eqref{Thresh_Z}, $\nabla g(\bZ^{(k)})$, and $\bY^{(k+1)}=\cP_{+}\left(\bZ^{(k)}-\nabla g(\bZ^{(k)})\right)$.
  \item[Step $\mathbf{2^\prime} c$] Compute $t^{(k+1)}=\frac{1+\sqrt{1+4{t^{(k)}}^2}}{2}$.
  \item[Step $\mathbf{2^\prime} d$] Compute $\bZ^{(k+1)}=\bY^{(k+1)}+\frac{t^{(k)}-1}{t^{(k+1)}}(\bY^{(k+1)}+\bY^{(k)})$.
\end{description}
\end{algorithm*}

The following theorem given by \citet{CuiL16} shows that the convergence of APG algorithm has $\cO(1/k^2)$ complexity. Its proof rests on Part 2 of Proposition \ref{Prop1} and Theorem 4.4 of \citet{Beck09}, which is a straight extension from vectors to matrices.

\begin{theorem}
\label{Thm_Covg}
Assume $f$ is defined in \eqref{Obj_Grad} and $\{\bZ_k\}$ are generated by the APG algorithm. Then for any optimal solution $\bZ^*$ of $\min_{\bZ} f(\bZ)$, we have
\begin{equation*}
    f(\bZ_k)-f(\bZ^*)\leq \frac{2\|\bZ_0-\bZ^*\|_F^2}{(k+1)^2},\quad \forall k \geq 1,
\end{equation*}
where $\bZ_0$ is the initial value.
\end{theorem}

Then, we conclude this part by the following theorem which shows that the value of objective function is nonincreasing and its proof is given as Appendix part \ref{sect.B}.
\begin{theorem}
\label{Thm1}
If we solve the problem \eqref{eq.minPnlhood} by the proposed iterative procedure, then the value of objective function in each iteration is nonincreasing.
\end{theorem}
\begin{remark}\rm
Compared with ADMM, the APG algorithm has a theoretical guarantee for the iterate number before reaching the prescribed accuracy and is free of penalty parameter selection which will be needed for the ADMM \citep{CuiL16}. The APG algorithm serves as an intermediate step to ensure $\hbSigma_{\be}^{(k)}$ to be positive definite. And this may give additional computation burden to our algorithm. Thus in the implementation of our algorithm we are more likely to solve \eqref{approxObjnoC} first and its solution is given by
\begin{equation}\label{}
   \hbSigma_{\be}^{(k+1)}=\cS(\cM_n,t\lambda \bW),
\end{equation}
which can be efficiently computed. When $\hbSigma_{\be}^{(k+1)}$ is positive definite, the algorithm goes to next iteration; when $\hbSigma_{\be}^{(k+1)}$ is not positive definite, the APG algorithm will be used to give a positive definite estimate of $\bSigma_{\be}$.
\end{remark}

\section{Empirical study}
\label{sect.4}
Note that in our model \eqref{eq.minPnlhood} a tuning parameter $\lambda$ should be selected. In this paper we use the $K$-fold cross-validation to choose $\lambda$, which has been widely used in covariance matrix estimation and finds its theoretical support in \citep{Bick08C}. Let $\mathcal{A}$ be the index set of $n$ observations, and $\bS_{\by,\mathcal{A}}=|\mathcal{A}|^{-1}\sum_{t\in \mathcal{A}}(\by_t-\bar{\by})(\by_t-\bar{\by})^T$ be the sample covariance matrix given by validation data. The symbol $|\mathcal{A}|$ denotes the cardinality of set $\mathcal{A}$. Let $\hbLambda(\mathcal{A}^c,\lambda)$ and $\hbSigma_{\be}(\mathcal{A}^c,\lambda)$ be the estimated loading matrix and error covariance matrix by the training data in $\mathcal{A}^c$, the compliment of $\mathcal{A}$, with tuning parameter $\lambda$. Then we partition the data into $K$ subsets, denoted by its index sets $\mathcal{A}_1,\cdots,\mathcal{A}_K$, and choose the $\lambda_{cv}$ as follows
\begin{align*}
    &\lambda_{cv}=\mathrm{arg}\min_{\lambda}\frac{1}{K}\sum_{k=1}^{K}L\left(\hbLambda(\mathcal{A}_k^c,\lambda), \hbSigma_{\be}(\mathcal{A}_k^c,\lambda),\bS_{\by,\cA_k}\right),
\end{align*}
where $L(\hbLambda,\hbSigma_{\be},\bS_{\by})=\frac{1}{n}\logdet\left(\hbLambda\hbLambda^T+\hbSigma_{\be}\right)+ \frac{1}{n}\tr(\bS_{\by}(\hbLambda\hbLambda^T+\hbSigma_{\be})^{-1}).$

{Now we specify some issues in the implementation of our algorithm. For \eqref{approxObj} in \textbf{Step} $\mathbf{2^{'}}$, we set the projection depth $t=0.1$, $\delta=10^{-4}$, and $\mathrm{P}_{\lambda}(\bSigma_{\be})$ is chosen as the SCAD penalty. When \eqref{approxObj} is solved  via the APG algorithm, we take the same termination criterion used in \citep{CuiL16}. As we have shown that our algorithm has a nonincreasing property, we terminate the algorithm when the reduction of the value of objective function is less than $10^{-6}$. We take the consistent estimates of $\bSigma_{\be}$ and $\bLambda$ given by \citet{BaiLI16} as the initial values in our iterative procedure. Moreover, the nonconvexity of the objective function \eqref{eq.minPnlhood} can not guarantee a unique minima, and a conventional method is to try different initial values and choose the one with minimum function value as the final estimator. }

\subsection{Simulation}
\label{sect.4.1}
In our simulation, we take the synthetic example given by \citet{Bai16} with some modification on the error covariance matrix $\bSigma_{\be}$ to check the efficiency of our estimator. Three error covariance matrices with different sparsity patterns are considered. In each setting, we generated 200 independent data sets. All the computations are performed in Matlab 2014a on a PC Intel Core(TM) i5-6600 CPU, 3.30Ghz with RAM 4.00 GB.

The underling factor model is constructed as follows. Let the two factors $\left\{f_{1j},f_{2j}\right\}$ be independently generated from $\cN(0,1)$, and  the elements of $\bLambda$ be uniform on $[0,1]$. The covariance matrix of the idiosyncratic error term has the following three different structures.
\begin{description}
  \item[Model 1.](\emph{Banded matrix}) Let $\left\{\alpha_{ij}\right\}_{i\leq p,j\leq n}$ be generated from the standard normal distribution $\cN(0,1)$, and
\begin{align*}
  & e_{1j}= \alpha_{1j},\\
  & e_{2j}= \alpha_{2j}+a_1\alpha_{1j},\\
  & e_{3j}= \alpha_{3j}+a_2\alpha_{2j}+b_1\alpha_{1j},\\
  & e_{(i+1)j}=\alpha_{(i+1)j}+a_i\alpha_{ij}+b_{i-1}\alpha_{(i-1)j}+c_{i-2}\alpha_{(i-2)j},
\end{align*}
with $\left\{a_i,b_i,c_i\right\}_{i\leq p}$ being generated from $0.7\times\cN(0,1)$.
  \item[Model 2.](\emph{Approximately sparse matrix}) Let $\bSigma_{\be}=\alpha \bI_p+\bM$, where $\bM=[m_{ij}]$ with $m_{ij}=0.5^{|i-j|}$, and $\alpha$ is used to control the condition number of $\bSigma_{\be}$ equal to $p$.
  \item[Model 3.](\emph{Block diagonal  matrix}) The indices $1,\cdots,p$ are evenly divided into 5 groups. Then for each group we set the covariance matrix to be AR(0.6) which means the conditional covariance between any two random variables $e_i$ and $e_j$ is $0.6^{|i-j|}$.

\end{description}
Since we want to investigate the influence of the positivity of $\bSigma_{\be}$ on the estimation of $\bLambda$ and $\bff$, we use three different methods: (1) MMEM: the majorize-minimize EM algorithm proposed by \citet{Bai16}; (2) EMAPG: our proposed algorithm; (3) EPC: a two-step algorithm based on generalized PC method \citep{Choi12} to estimate the loading matrix and factor, and principal orthogonal complement thresholding (POET) method \citep{FanL13} to estimate the error covariance matrix. When using POET estimator, we follow \citet{Bai16} to choose the thresholding parameter $C=0.7,0.5$.

{We compare the performance of these three different estimation procedures with respect to the following aspects. For the factor loading matrix (LOAD) and the factors (FACT), we take two different criterions. One is the smallest canonical correlation (CCOR) between the estimator and the parameter (the higher the better), which was also used in \citep{Bai16} as a measurement to assess the accuracy of each estimator; the other is mean squared error (MSE), and smaller MSE means better accuarcy. For the error covariance matrix, we report the root-mean-square error(RMSE)
\begin{align*}
\mathrm{RMSE}_{\bSigma_{\be}}=\frac{1}{200\times p}\sum_{t=1}^{200} \left\|\hbSigma_{\be}(t)-\bSigma_{\be}\right\|_F
\end{align*}
and the empirical deviation (in parentheses) across 200 replications. As an easy result with \eqref{decomSgmy}, the RMSE of $\bSigma_{\by}$ denoted by $\mathrm{RMSE}_{\bSigma_{\by}}$ is also reported. Moreover, we also exhibit the ratios of $\hbSigma_{\be}$ and $\hbSigma_{\by}$ being non-positive definite (RNPD) in $200$ replications. Larger ratio means worse approximation of the underling covariance structure. All the simulation results for Models 1, 2 and 3 are reported in Tables \ref{Table1}, \ref{Table2}, \ref{Table3}, \ref{Table4}, \ref{Table31} and \ref{Table32}, correspondingly.}

From Tables \ref{Table1}, \ref{Table3} and \ref{Table31}, we note that except the MSEs of loading matrix for EMAPG and MMEM algorithms are similar, our EMAPG algorithm outperforms the MMEM algorithm with respect to CCOR and MSE, which implies that preserving the positive definiteness of $\hbSigma_{\be}$ can improve the penalized MLE of $\bLambda$ and $\bff$. And from Tables \ref{Table2}, \ref{Table4} and \ref{Table32}, we also note that the positive definiteness of $\hbSigma_{\be}$ plays an important role in the WLS estimation of $\bff$, and when RNPD of $\hbSigma_{\be}$ (or $\hbSigma_{\by}$) is higher, the superiority of our algorithm becomes more obvious. Moreover, as a byproduct the EMAPG algorithm also has very good performance in estimating the covariance structure of the approximate factor model. From Tables \ref{Table3} and \ref{Table31}, we note that the EPC algorithm may give best performance when $\hbSigma_{\be}$ is positive definite. This coincides with the conclusion given in \citep{Bai16} that it is hard to see whether MMEM(EMAPG) or EPC dominates the other, but no matter which method will be used a positive definite estimate of $\bSigma_{\be}$ is essential to the WLS estimate of $\bff$. We note that the EPC method also has very good performance in estimating the covariance structure of approximate factor model with respect to RMSE. But, due to the thresholding procedure in POET, the EPC method can not guarantee the positive definiteness of $\hbSigma_{\be}$, and becomes inefficiency in estimating $\bff$. We may conclude that our estimator \eqref{eq.minPnlhood} not only gives simultaneous estimates of the factor loading matrix $\bLambda$ and the error covariance matrix $\bSigma_{\be}$, but also can be used to estimate the covariance structure of the approximate factor model. According to our simulation, keeping the positive definiteness of the error covariance matrix can improve the estimation of loading matrix and factors (with WLS method) for both the one step penalized MLE and the two step EPC methods. {Another issue is CPU time. The the iterative step in keeping $\hbSigma_{\be}$ positive definite brings much more computational burden to EMAPG algorithm compared with MMEM algorithm. But the EMAPG algorithm does give best performance in our simulation.}
\begin{table*}[htp]
\centering{
  \caption[1-1]{Comparison of three methods with respect to loading matrix (LOAD) and factors(FACT). \label{Table1}}
  \begin{tabular}{lllrrrrrrrr}
 \midrule
  \textbf{Model 1}&&&\multicolumn{2}{c}{MMEM}&\multicolumn{2}{c}{EMAPG}&\multicolumn{4}{c}{EPC}  \\
  \cline{8-11}
            &     &       & \multicolumn{2}{c}{$\lambda_{CV}$} &\multicolumn{2}{c}{$\lambda_{CV}$}   & \multicolumn{2}{c}{$C=0.7$}   &  \multicolumn{2}{c}{$C=0.5$} \\
  \cline{4-11}
            & $n$ &  $p$ & CCOR&$\MSE$& CCOR  &$\MSE$&  CCOR  &$\MSE$ & CCOR  &$\MSE$\\
  \hline
  LOAD      & 50  & 50  &0.3954 &0.0425 &\bf{0.4043} &0.0428 &0.3430 &0.0441 &0.3278 &0.0406\\
            &     & 100 &0.4940 &0.0572 &\bf{0.5529} &0.0578 &0.4168 &0.0616 &0.3604 &0.0573\\
            &     & 150 &0.6111 &0.0681 &\bf{0.6407} &0.0688 &0.5967 &0.0701 &0.4792 &0.0686\\
            &     & 200 &0.6457 &0.0770 &\bf{0.6857} &0.0779 &0.6741 &0.0805 &0.5310 &0.0741\\
  \cline{2-11}
            & 100 & 50  &0.4704 &0.0390 &\bf{0.4876} &0.0398 &0.4612 &0.0419 &0.4175 &0.0409\\
            &     & 100 &0.5980 &0.0544 &\bf{0.6621} &0.0552 &0.4686 &0.0577 &0.3895 &0.0562\\
            &     & 150 &0.7917 &0.0696 &\bf{0.8146} &0.0705 &0.7767 &0.0710 &0.6307 &0.0679\\
            &     & 200 &0.8138 &0.0732 &\bf{0.8301} &0.0746 &0.8018 &0.0765 &0.6428 &0.0719\\
  \hline
  FACT      & 50  & 50  &0.3833 &0.2552 &\bf{0.3936} &0.0728 &0.3522 &0.0711 &0.3330 &0.0680\\
            &     & 100 &0.6070 &0.0876 &\bf{0.6957} &0.0692 &0.5426 &0.0698 &0.4308 &0.0685\\
            &     & 150 &0.7298 &0.0752 &\bf{0.7929} &0.0693 &0.7490 &0.0685 &0.5195 &0.0708\\
            &     & 200 &0.7909 &0.1000 &\bf{0.8416} &0.0676 &0.8358 &0.0686 &0.5680 &0.0694\\
  \cline{2-11}
            & 100 & 50  &0.4457 &0.1301 &\bf{0.4608} &0.1054 &0.4251 &0.0981 &0.3731 &0.0983\\
            &     & 100 &0.6547 &0.1183 &\bf{0.7186} &0.1002 &0.5271 &0.0975 &0.3663 &0.0996\\
            &     & 150 &0.8502 &0.1489 &\bf{0.8998} &0.0970 &0.8736 &0.0938 &0.5744 &0.0971\\
            &     & 200 &0.8620 &0.1201 &\bf{0.9052} &0.0965 &0.8581 &0.0949 &0.5640 &0.0965\\
\bottomrule
\end{tabular}}
\end{table*}

\begin{landscape}
\begin{table*}[htp]
\centering{
  \caption{Comparison of three methods with respect to $\bSigma_{\be}$ and $\bSigma_{\by}$. \label{Table2}}
  \begin{tabular}{lcccccc}
 \midrule
  \textbf{Model 1}&      &    &  MMEM  &  EMAPG  & \multicolumn{2}{c}{EPC}  \\
  \cline{6-7}
            &     &      & $\lambda_{CV}$ &$\lambda_{CV}$ & $C=0.7$ & $C=0.5$ \\
  \hline
            &     &      & ${\bSigma_{\be}}\mid{\bSigma_{\by}}$& ${\bSigma_{\be}}\mid{\bSigma_{\by}}$ & ${\bSigma_{\be}}\mid{\bSigma_{\by}}$ & ${\bSigma_{\be}}\mid{\bSigma_{\by}}$\\
  \cline{4-7}
  RNPD      & 50  & 50   & 0.0850$\mid$0.0350 & 0$\mid$0 & 0$\mid$0 & 0.1600$\mid$0.1600   \\
            &     & 100  & 0.3150$\mid$0.2800 & 0$\mid$0 & 0$\mid$0 & 0.8850$\mid$0.8850    \\
            &     & 150  & 0.6450$\mid$0.5550 & 0$\mid$0 & 0$\mid$0 & 1$\mid$1 \\
            &     & 200  & 0.7800$\mid$0.7600 & 0$\mid$0 & 0$\mid$0 & 1$\mid$1  \\
  \cline{2-7}
            & 100 & 50   &0.1850$\mid$0.0050  &0$\mid$0 &  0$\mid$0        & 0.4350$\mid$0.4350 \\
            &     & 100  &0.1950$\mid$0.1100  &0$\mid$0 &0.0800$\mid$0.0800&  0.9950$\mid$0.9950\\
            &     & 150  &0.4950$\mid$0.4300  &0$\mid$0 &0.0050$\mid$0.0050&  1$\mid$1 \\
            &     & 200  &0.5950$\mid$0.5150  &0$\mid$0 &0.4950$\mid$0.4950&  1$\mid$1 \\
  \hline
  $\mathrm{RMSE}_{\bSigma_{\be}}$&
                50& 50   &4.4333(3.0957e+03)    &0.2781(7.5245e-04) &0.2544(1.4590e-04) & 0.2406(1.9101e-04)\\
            &     & 100  &28.2653(1.4710+05)    &0.2022(3.3704e-05) &0.1717(3.5727e-05) & 0.1687(4.2260e-05)\\
            &     & 150  &1.3006(61.4565)       &0.1540(3.7551e-05) &0.1212(1.4117e-05) & 0.1261(1.3933e-05)\\
            &     & 200  &2.4525e+03(1.0820e+09)&0.1439(6.2364e-05) &0.1206(7.9970e-06) & 0.1308(9.5932e-06) \\
  \cline{2-7}
            & 100 & 50   &1.8999(201.9690)    &0.3014(7.5342e-04) &0.2285(1.2228e-04) &0.2125(1.4658e-04) \\
            &     & 100  &1.7287(89.3294)     &0.1845(4.5147e-05) &0.1364(6.4340e-05) &0.1258(8.1845e-05) \\
            &     & 150  &8.8185(5.3643e+03)  &0.1455(4.1528e-05) &0.0981(2.9208e-05) &0.0942(3.4040e-05) \\
            &     & 200  &6.8095(4.9002e+03)  &0.1382(1.5709e-05) &0.0891(7.0492e-06) &0.0891(5.4020e-06) \\
  \hline
  $\mathrm{RMSE}_{\bSigma_{\by}}$&
              50  & 50   &4.5434(3.0948e+03)    &0.3934(0.0021) &0.4256(0.0022) &0.4107(0.0024) \\
            &     & 100  &28.4198(1.4709e+05)   &0.3769(0.0016) &0.3805(0.0018) &0.3779(0.0040) \\
            &     & 150  &1.4485(61.1368)       &0.3295(0.0015) &0.3254(0.0015) &0.3411(0.0050) \\
            &     & 200  &2.4526e+03(1.0820e+09)&0.3328(0.0011) &0.3263(0.0013) &0.3506(0.0056) \\
  \cline{2-7}
            & 100 & 50   &1.9580(201.7909) &0.3436(0.3436)    &0.3417(0.0016)     &0.3166(0.0014) \\
            &     & 100  &1.8136(89.0758)     &0.2764(7.1768e-04)&0.2585(9.2646e-04) &0.2623(0.0049) \\
            &     & 150  &8.9094(5.3627e+03)  &0.2548(5.3078e-04)&0.2366(7.2468e-04) &0.2710(0.0089) \\
            &     & 200  &6.9032(4.8990e+03)  &0.2500(5.0955e-04)&0.2417(0.0027)     &0.2701(0.0070) \\
\bottomrule
\end{tabular}}
\end{table*}
\end{landscape}

\begin{table*}[htp]
\centering{
  \caption{Comparison of three methods with respect to loading matrix (LOAD) and factors(FACT). \label{Table3}}
  \begin{tabular}{lllrrrrrrrr}
 \midrule
  \textbf{Model 2}&&&\multicolumn{2}{c}{MMEM}&\multicolumn{2}{c}{EMAPG}&\multicolumn{4}{c}{EPC}  \\
  \cline{8-11}
            &     &       & \multicolumn{2}{c}{$\lambda_{CV}$} &\multicolumn{2}{c}{$\lambda_{CV}$}   & \multicolumn{2}{c}{$C=0.7$}   &  \multicolumn{2}{c}{$C=0.5$} \\
  \cline{4-11}
            & $n$ &  $p$ & CCOR&$\MSE$& CCOR  &$\MSE$&  CCOR  &$\MSE$ & CCOR  &$\MSE$\\
  \hline
  LOAD      & 50  & 50  &0.7825 &0.0410 &0.8316      &0.0415 &\bf{0.8535} &0.0412 &0.7987 &0.0396\\
            &     & 100 &0.7892 &0.0521 &0.8566      &0.0526 &\bf{0.8764} &0.0569 &0.7724 &0.0522\\
            &     & 150 &0.8496 &0.0668 &\bf{0.8938} &0.0678 &0.8744      &0.0685 &0.7633 &0.0635\\
            &     & 200 &0.8732 &0.0767 &\bf{0.9006} &0.0779 &0.8533      &0.0774 &0.6964 &0.0735\\
  \cline{2-11}
            & 100 & 50  &0.8732 &0.0395 &0.9074      &0.0403 &\bf{0.9205} &0.0402 &0.8160 &0.0387\\
            &     & 100 &0.8796 &0.0542 &\bf{0.9365} &0.0545 &0.7843      &0.0489 &0.8242  &0.0516\\
            &     & 150 &0.8788 &0.1284 &\bf{0.9425} &0.1278 &0.7206      &0.1228 &0.8110 &0.1217\\
            &     & 200 &0.8991 &0.0755 &\bf{0.9501} &0.0745 &0.7072      &0.0711 &0.8034 &0.0703\\
  \hline
  FACT      & 50  & 50  &0.7215 &0.2033 &0.8464      &0.0717 &\bf{0.8889} &0.0688 &0.7018 &0.0688\\
            &     & 100 &0.7063 &1.9261 &0.9061      &0.0694 &\bf{0.9521} &0.0697 &0.6514 &0.0699\\
            &     & 150 &0.7275 &0.3294 &\bf{0.9515} &0.0697 &0.8879      &0.0683 &0.6117 &0.0685\\
            &     & 200 &0.7405 &0.1254 &\bf{0.9647} &0.0682 &0.8256      &0.0701 &0.5403 &0.0693\\
  \cline{2-11}
            & 100 & 50  &0.7107 &0.2797 &0.8684      &0.0985 &\bf{0.9095} &0.0953 &0.5776 &0.0970\\
            &     & 100 &0.7172 &0.3958 &\bf{0.9489} &0.1004 &0.5371      &0.0949 &0.6040 &0.0976\\
            &     & 150 &0.6986 &44.5154&\bf{0.9729} &0.2041 &0.4748      &0.1980 &0.5698 &0.1997\\
            &     & 200 &0.6816 &1.4890 &\bf{0.9850} &0.0983 &0.4411      &0.0981 &0.5845 &0.0968\\
\bottomrule
\end{tabular}}
\end{table*}

\begin{landscape}
\begin{table*}[htp]
\centering{
  \caption{Comparison of three methods with respect to $\bSigma_{\be}$ and $\bSigma_{\by}$. \label{Table4}}
  \begin{tabular}{lcccccc}
 \midrule
  \textbf{Model 2}&      &    &  MMEM  &  EMAPG  & \multicolumn{2}{c}{EPC}  \\
  \cline{6-7}
            &     &      & $\lambda_{CV}$ &$\lambda_{CV}$ & $C=0.7$ & $C=0.5$ \\
  \hline
            &     &      & ${\bSigma_{\be}}\mid{\bSigma_{\by}}$& ${\bSigma_{\be}}\mid{\bSigma_{\by}}$ & ${\bSigma_{\be}}\mid{\bSigma_{\by}}$ & ${\bSigma_{\be}}\mid{\bSigma_{\by}}$\\
  \cline{4-7}
  RNPD      & 50  & 50   &0.9550$\mid$0.9350 & 0$\mid$0 &    0$\mid$0      &  1$\mid$1 \\
            &     & 100  &0.9950$\mid$0.9950 & 0$\mid$0 &    0$\mid$0      &  1$\mid$1 \\
            &     & 150  &  1$\mid$1         & 0$\mid$0 &0.8050$\mid$0.8050&  1$\mid$1 \\
            &     & 200  &0.9950$\mid$0.9900 & 0$\mid$0 &      1$\mid$1    &  1$\mid$1 \\
  \cline{2-7}
            & 100 & 50   &0.9650$\mid$0.9600& 0$\mid$0 &0.0250$\mid$0.0050& 1$\mid$1  \\
            &     & 100  &  1$\mid$0.9950   & 0$\mid$0 &   1$\mid$1       & 1$\mid$1  \\
            &     & 150  &0.9900$\mid$0.9900& 0$\mid$0 &   1$\mid$1       & 1$\mid$1  \\
            &     & 200  &  1$\mid$1        & 0$\mid$0 &   1$\mid$1       &  1$\mid$1 \\
  \hline
$\mathrm{RMSE}_{\bSigma_{\be}}$&
              50 & 50   &386.2803(1.8140e+07)  &0.1110(3.8297e-06) &0.0836(1.5029e-05)&0.0752(1.8411e-05) \\
            &    & 100  &9.5761e+04(1.8308e+12)&0.0785(2.5924e-06) &0.0519(3.4325e-06)&0.0484(2.9666e-06) \\
            &    & 150  &1.0811e+04(2.2867e+10)&0.0645(2.5259e-06) &0.0403(1.3453e-06)& 0.0398(1.2129e-06 \\
            &    & 200  &75.7420(2.2395e+05)   &0.0573(1.2244e-06) &0.0344(7.4460e-07)&0.0356(5.8684e-07) \\
  \cline{2-7}
            &100 & 50   &53.2334(2.3425e+05)   &0.1091(2.2362e-06) &0.0705(7.5454e-06) &0.0627(7.3771e-06)  \\
            &    & 100  &446.8518(2.2263e+07)  &0.0766(2.0965e-06) &0.0399(2.5089e-06) &0.0366(2.0241e-06)  \\
            &    & 150  &1.7385e+05(5.6536e+12)&0.0623(1.0231e-06) &0.0306(8.5315e-07) &0.0294(7.2741e-07) \\
            &    & 200  &323.6460(5.8833e+06)  &0.0536(5.8271e-07) &0.0257(4.7981e-07) &0.0258(4.5756e-07)  \\
  \hline
$\mathrm{RMSE}_{\bSigma_{\by}}$&
              50  & 50   &386.3244(1.8140e+07)  &0.2235(0.0023) &0.2223(0.0037)  &0.2390(0.0076) \\
            &     & 100  &9.5762e+04(1.8308e+12)&0.1954(0.0019) &0.1897(0.0028)  &0.2374(0.0093)  \\
            &     & 150  &1.0811e+04(2.2867e+10)&0.1889(0.0016) & 0.1946(0.0037) &0.2355(0.0094)  \\
            &     & 200  &75.7767(2.2394e+05)   &0.1899(0.0022) &0.2108(0.0075)  & 0.2690(0.0145)  \\
  \cline{2-7}
            & 100 & 50   &53.2582(2.3425e+05)   &0.1741(0.0010) &0.1567(0.0014) &0.2109(0.0083)  \\
            &     & 100  &446.8674(2.2263e+07)  &0.1593(0.0015) &0.2263(0.0158) &0.2149(0.0152) \\
            &     & 150  &1.7385e+05(5.6536e+12)&0.1555(0.0016) &0.2459(0.0148) &0.2295(0.0158)  \\
            &     & 200  &323.6604(5.8833e+06)  &0.1580(0.0021) &0.2251(0.0120) &0.2280(0.0161) \\
\bottomrule
\end{tabular}}
\end{table*}
\end{landscape}


\begin{table*}[htp]
\centering{
  \caption{Comparison of three methods with respect to loading matrix (LOAD) and factors(FACT). \label{Table31}}
  \begin{tabular}{lllrrrrrrrr}
 \midrule
  \textbf{Model 3}&&&\multicolumn{2}{c}{MMEM}&\multicolumn{2}{c}{EMAPG}&\multicolumn{4}{c}{EPC}  \\
  \cline{8-11}
            &     &       & \multicolumn{2}{c}{$\lambda_{CV}$} &\multicolumn{2}{c}{$\lambda_{CV}$}   & \multicolumn{2}{c}{$C=0.7$}   &  \multicolumn{2}{c}{$C=0.5$} \\
  \cline{4-11}
            & $n$ &  $p$ & CCOR&$\MSE$& CCOR  &$\MSE$&  CCOR  &$\MSE$ & CCOR  &$\MSE$\\
  \hline
  LOAD      & 50  & 50  & 0.5922 & 0.0407 & 0.6071 & 0.0411 &\bf{0.6377} & 0.0421 & 0.6454 & 0.0433\\
            &     & 100 & 0.7463 & 0.0542 & 0.7856 & 0.0549 &\bf{0.7943} & 0.0536 & 0.7351 & 0.0546\\
            &     & 150 & 0.8165 & 0.0700 & 0.8540 & 0.0711 &\bf{0.8593} & 0.0695 & 0.7669 & 0.0661\\
            &     & 200 & 0.8171 & 0.0788 & 0.8480 & 0.0802 &\bf{0.8554} & 0.0793 & 0.6941 & 0.0761\\
  \cline{2-11}
            & 100 & 50  & 0.6301 & 0.0373 & \bf{0.6490} & 0.0376 &  0.6307      & 0.0374 & 0.6344 & 0.0375\\
            &     & 100 & 0.8647 & 0.0553 & 0.8854      & 0.0563 &  \bf{0.8897} & 0.0530 & 0.8280 & 0.0527\\
            &     & 150 & 0.8991 & 0.0690 & 0.9193      & 0.0701 &  \bf{0.9215} & 0.0673 & 0.8452 & 0.0645\\
            &     & 200 & 0.9028 & 0.0731 & 0.9242      & 0.0736 &  \bf{0.9276} & 0.0716 & 0.8442 & 0.0679\\
  \hline
  FACT      & 50  & 50  & 0.5580 & 0.3452 & 0.6141 & 0.0690 &\bf{0.6521} & 0.0678 & 0.6624 & 0.0704\\
            &     & 100 & 0.7674 & 0.1090 & 0.8617 & 0.0703 &\bf{0.8731} & 0.0667 & 0.6969 & 0.0692\\
            &     & 150 & 0.8162 & 0.5846 & 0.9344 & 0.0698 &\bf{0.9462} & 0.0668 & 0.6934 & 0.0682\\
            &     & 200 & 0.8478 & 0.3853 & 0.9397 & 0.0689 &\bf{0.9528} & 0.0686 & 0.6008 & 0.0713\\
  \cline{2-11}
            & 100 & 50  & 0.5798 & 0.1833 &\bf{0.6185} & 0.1012 &  0.6056      & 0.0970 & 0.6090 & 0.0968\\
            &     & 100 & 0.7423 & 0.1620 & 0.8946     & 0.1006 & \bf{0.9068}  & 0.0936 & 0.7262 & 0.0968\\
            &     & 150 & 0.8373 & 0.2067 & 0.9703     & 0.1004 &  \bf{0.9753} & 0.0952 & 0.7224 & 0.0973\\
            &     & 200 & 0.8146 & 0.2241 & 0.9709     & 0.0958 &  \bf{0.9765} & 0.0927 & 0.6896 & 0.0947\\
\bottomrule
\end{tabular}}
\end{table*}

\begin{landscape}
\begin{table*}[htp]
\centering{
  \caption{Comparison of three methods with respect to $\bSigma_{\be}$ and $\bSigma_{\by}$. \label{Table32}}
  \begin{tabular}{lcccccc}
 \midrule
  \textbf{Model 3}&      &    &  MMEM  &  EMAPG  & \multicolumn{2}{c}{EPC}  \\
  \cline{6-7}
            &     &      & $\lambda_{CV}$ &$\lambda_{CV}$ & $C=0.7$ & $C=0.5$ \\
  \hline
            &     &      & ${\bSigma_{\be}}\mid{\bSigma_{\by}}$& ${\bSigma_{\be}}\mid{\bSigma_{\by}}$ & ${\bSigma_{\be}}\mid{\bSigma_{\by}}$ & ${\bSigma_{\be}}\mid{\bSigma_{\by}}$\\
  \cline{4-7}
  RNPD      & 50  & 50   & 0.8450$\mid$0.8200 & 0$\mid$0 &   0$\mid$0 & 0.0050$\mid$0.0050 \\
            &     & 100  & 0.9850$\mid$0.9750 & 0$\mid$0 &   0$\mid$0 & 0.9450$\mid$0.9450 \\
            &     & 150  &  1$\mid$0.9950     & 0$\mid$0 &   0$\mid$0 &   1$\mid$1   \\
            &     & 200  & 0.9850$\mid$0.9850 & 0$\mid$0 &   0$\mid$0 &   1$\mid$1   \\
  \cline{2-7}
            & 100 & 50   & 0.2100$\mid$0.2000 &   0$\mid$0   &   0$\mid$0   &  0.1800$\mid$0.1800 \\
            &     & 100  & 0.9150$\mid$0.8800 &   0$\mid$0   &   0$\mid$0   & 0.5000$\mid$0.5000 \\
            &     & 150  &   1$\mid$0.9900    &   0$\mid$0   &   0$\mid$0   &   0$\mid$0   \\
            &     & 200  & 0.9900$\mid$0.9800 &   0$\mid$0   &   0$\mid$0   &   1$\mid$1   \\
  \hline
$\mathrm{RMSE}_{\bSigma_{\be}}$&
              50 & 50   & 17.6165(8.5831e+03)  & 0.1342(6.8154e-06) & 0.1211(1.5340e-05)& 0.1080(3.0385e-05) \\
            &    & 100  & 302.9665(1.3792e+07)  & 0.0980(4.0206e-06) & 0.0810(5.1237e-06)& 0.0723(7.0484e-06) \\
            &    & 150  & 1.3785e+04(2.8936e+10)  &  0.0817(3.9691e-06) & 0.0653(2.3473e-06)& 0.0609(2.3916e-06) \\
            &    & 200  & 4.3840e+04(3.8321e+11)  & 0.0716(2.9771e-06) & 0.0565(1.2709e-06)& 0.0550(1.1393e-06) \\
  \cline{2-7}
            &100 & 50   & 3.7862e+03(2.8564e+09)  & 0.1276(2.6228e-06) & 0.1094(1.1661e-05)&  0.0931(2.2638e-05) \\
            &    & 100  & 33.3834(9.6090e+04)  & 0.0931(1.0097e-06) & 0.0682(4.4114e-06)& 0.0558(4.2159e-06) \\
            &    & 150  & 238.4404(6.6527e+06)  & 0.0769(4.7543e-07) & 0.0525(2.1810e-06)& 0.0444(1.5492e-06) \\
            &    & 200  & 80.4486(9.5889e+05)  & 0.0670(3.3169e-07) & 0.0451(1.0238e-06)& 0.0396(6.5967e-07) \\
  \hline
$\mathrm{RMSE}_{\bSigma_{\by}}$&
              50  & 50   & 17.6840(8.5807e+03)  & 0.2690(0.0014) & 0.2246(0.0016)& 0.2177(0.0017) \\
            &     & 100  & 303.0102(1.3792e+07)  & 0.2303(0.0016) & 0.2268(0.0018)& 0.2363(0.0046) \\
            &     & 150  & 1.3785e+04(2.8936e+10) & 0.2308(0.0019) &  0.2267(0.0021)& 0.0021(0.0053) \\
            &     & 200  & 4.3841e+04(3.8321e+11)  & 0.2205(0.0021) & 0.2176(0.0023)& 0.2484(0.0070) \\
  \cline{2-7}
            & 100 & 50   & 3.7862e+03(2.8564e+09)  & 0.1845(6.3202e-04) & 0.1808(9.2587e-04)& 0.1674(9.9825e-04) \\
            &     & 100  & 33.4172(9.6088e+04)  & 0.1741(0.0011) & 0.1663(0.0014)& 0.1674(0.0025) \\
            &     & 150  & 238.4683(6.6527e+06)  & 0.1653(8.9474e-04) & 0.1581(0.0012)& 0.1847(0.0047) \\
            &     & 200  & 80.4751(9.5888e+05)  & 0.1522(6.6468e-04) & 0.1464(7.8750e-04)& 0.1753(0.0043) \\
\bottomrule
\end{tabular}}
\end{table*}
\end{landscape}

\subsection{Forecasting simulation}
\label{sect.4.2}
The difference between ML based and PC based methods for estimating the high-dimensional approximate factor model has been extensively discussed \citep{Bai16}. \citet{Bai16} also showed that incorporating the cross-sectional correlations does lead to a considerable improvement on the estimation of $\bLambda$ and $\bff$ and a better performance of forecasting. In our simulation, we have shown that a positive definite estimate of $\bSigma_{\be}$ can further improve the estimation of $\bLambda$ and $\bff$, and now we will check its influence on forecasting through simulation.

Within a same framework, we only consider the following three ML based methods to investigate the influence of different estimators of $\bSigma_{\be}$ on forecasting: (1) the heteroscedastic ML (HML) estimator \citep{BaiLI16} that gives no concern on the cross-sectional dependence and estimates $\bSigma_{\be}$ to be diagonal; (2) the MMEM algorithm that takes the cross-sectional dependence into consideration but can not guarantee $\hbSigma_{\be}$ to be positive definite; (3) the EMAPG algorithm that not only takes the cross-sectional dependence into consideration but also guarantees $\hbSigma_{\be}$ to be positive definite. We construct the following synthetic two factors time series model:
\begin{align}
\label{simulFmodl}
    & x_{t+1}=\bbeta^T\bff_t+\epsilon_t,\quad \bff_{t}=\rho \bff_{t-1}+\bnu_t,
\end{align}
where $\bbeta=[2, 3 ]^T$, $\rho=0.5$, $\epsilon_t \sim \cN(0,1)$ and $\bnu_t\sim \cN(0,I_2)$. The factor $\bff_t$ in \eqref{simulFmodl} is unknown and can be estimated from the following factor model
\begin{equation*}
    \by_t=\bLambda \bff_t+\be_t,
\end{equation*}
where the elements in $\bLambda$ are generated from the uniform distribution on the interval $[0,1]$, the components of $\bff_t$ are generated from standard normal distribution, and $\be_t$ has the same covariance structure as Model 2.

In our simulation we set $n=50,100$. To compare the forecasting error, we generate $m+n$ observations, and then conduct one-step ahead out-of-sample forecasting $m$ times with a moving window of a fixed sample size. Specially, setting $t=0,\cdots,m-1$, we first estimate $\bff$ by the data $\{\by_{i}\}_{i=t+1}^{t+n}$ and get $\{\hat{\bff}_i\}_{i=t+1}^{t+n}$, then the estimator $\hat{\bbeta}_{t+n}$ is obtained by regressing $\{x_{i}\}_{i=t+2}^{t+n}$ onto $\{\hat{\bff}_{i}\}_{i=t+1}^{t+n-1}$. The forecast of $x_{t+n+1}$ is $\hat{x}_{t+n+1}=\hat{\bbeta}_{t+n}^T\hat{\bff}_{t+n}$, and the forecasting error is given by $(x_{t+n+1}-\hat{x}_{t+n+1})^2$. In order to give quantitative measurement of the performance, we take the PC estimator as benchmark and compute the mean squared out-of-sample forecasting error by
\begin{equation*}
\mathrm{MSE}=\frac{1}{m}\sum_{t=0}^{m-1}(x_{t+T+1}-\hat{x}_{t+T+1})^2.
\end{equation*}
For different methods, the relative MSEs to PC method are reported in Table \ref{Table5}.
\begin{table}[htp]
\centering
  \caption{Forecasting error of synthetic data with $m=50$. \label{Table5}}
  \begin{tabular}{llccc}
    \hline
          &         & HML    & MMEM    & EMAPG \\
  \hline
   $n=50$ & $p=50$   & 0.8411 & 0.7683 & 0.6803\\
          & $p=100$  & 0.9291 & 0.9241 & 0.8738\\
          & $p=150$  & 0.9583 & 0.9595 & 0.8993\\
  \hline
   $n=100$& $p=50$   &0.8558 & 0.7881 & 0.7549\\
          & $p=100$  &0.9441 & 1.0515 & 0.9851\\
          & $p=150$  &0.9965 & 1.0085 & 0.9992\\
    \hline
  \end{tabular}
\end{table}
From Table \ref{Table5}, we find that considering the correlation between idiosyncratic error term can improve the performance of forecasting, and the MSEs of EMAPG algorithm  are usually smaller than the other two methods, which implies that the forecasting accuracy can be further improved by preserving the positive definiteness of $\hbSigma_{\be}$. Moreover, we can also find that as $n$ and $p$ increase  the performances of these three methods become very similar, which coincides with the consistency properties given in \citep{Bai16,BaiLI16}.

\subsection{Macroeconomic times series data}

The diffusion index forecasting model \citep{Stock02} has the following form
\begin{align*}
  x_{t+h}^h& = a_h+\bbeta^T_h \bff_t+\sum_{i=1}^{l}\gamma_{ih}x_{t+1-i}+\epsilon_{t+h}^{h}, \\ \by_t &= \bLambda \bff_t+\be_t ,
\end{align*}
where $h$ is the forecast horizon, $l$ is the number of lags for $x_t$, and the goal is to forecast $x_{t+h}^h=\frac{1}{h}\sum_{i=1}^hx_{t+i}$, the $h$-step-ahead variable. Here we forecast the industrial production with the dataset of real time macroeconomic time series of the United States, which has been analyzed by \citet{Ludv11} and identified with 8 factors by information criterion. We take the same procedure of forecasting used in Section \ref{sect.4.2} and report the relative MSEs to PC method in Table \ref{Table6}.
\begin{table}[htp]
\centering
  \caption{Forecasting error of real data analysis. \label{Table6}}
  \begin{tabular}{llrrr}
    \hline
           &           &HML     & MMEM   & EMAPG \\
    \hline
    $h=1$  &  One leg  &  1.0696& 1.0389 & 1.0528 \\
           &  Three leg&  1.0696& 1.0264 & 1.0265 \\
   $h=9 $  &  One leg  &  1.1320& 0.9152 & 0.9044\\
           &  Three leg&  1.1303& 0.9156 & 0.8679\\
   $h=18$  &  One leg  &  0.9426& 0.8316 & 0.7154\\
           &  Three leg&  0.9387& 0.8270 & 0.6864\\
    \hline
  \end{tabular}
\end{table}

From Table \ref{Table6}, when $h=1$, there are very little differences among these three methods on short term forecasting, and we can not tell which method should be preferred. The same findings have been illustrated by \citep{Bai16,BoNg05,Luci14}. We note that taking the cross-sectional dependence into account makes MMEM and EMAPG have smaller MSEs compared with HML method for long term forecasting. Moreover, our numerical results also show that the MSEs of EMAPG algorithm are smaller than that of MMEM for the cases $h=9$ and $h=18$, which means that preserving the positive definiteness of $\hbSigma_{\be}$ can improve the performance of forecasting. On the other side, the reliability of long term forecasting may suffer from the potential loss of the stationarity \citep{Bai16}.

\section{Concluding discussion}
\label{sect.5}
In this paper, by revisiting the penalized MLE of the high-dimensional approximate factor model proposed by \citet{Bai16}, we propose a new estimator and present the corresponding algorithm. The main contribution of our work is that we reformulate the estimation of error covariance matrix as a high-dimensional covariance matrix estimation problem, and propose an APG algorithm to guarantee the estimate of error covariance matrix to be positive definite. Combined with the APG algorithm, an EM iterative procedure (EMAPG algorithm) is given to give simultaneous estimates of the factor loading matrix and the error covariance matrix. We also prove that although the EMAPG algorithm can not guarantee a global minimizer, it enjoys a desirable non-increasing property. Our simulation shows that the proposed EMAPG algorithm not only gives a positive definite estimation of error covariance matrix but also improves the estimation of the high-dimensional approximate factor model. Moreover, we note that a positive definite estimate of error covariance matrix also improves the performance of forecasting when the factors are estimated by the WLS method. As an interesting byproduct, our estimator can be easily used to estimate the covariance matrix of the approximate factor model, which does not hold for the estimator given in \citep{Bai16}. On the other side, keeping the estimate of error covariance matrix positive definite in every iteration may make the EMAPG algorithm time  consuming, thus finding some easy computable and positive definite estimates of the error covariance matrix to reduce the computational burden should be treated as a research topic in the near future.

\section*{Acknowledgement}The authors wish to thank the anonymous reviewers and the Editor for their helpful and very detailed comments, which have led a substantial improvement to the presentation of their paper.  They also appreciate Yuan Liao for sharing the code used in \citet{Bai16}.

\appendix
\section*{Appendix}
\section{The Moreau-Yosida regularization}
\label{MYR}
The Moreau-Yosida regularization of a closed proper convex function $f$ associated with a given parameter $\rho$ is defined as
\begin{equation}\label{M-Yregu}
    \varphi_{f}^{\rho}(x)=\min_{y\in \cX}\left\{f(y)+\frac{1}{2\rho}\|y-x\|^2\right\},
\end{equation}
where $x\in \cX$, the domain of $f$ and equipped with the norm $\|\cdot\|$. The unique minimizer of $\eqref{M-Yregu}$, denoted by $P_{f}^{\rho}(x)$, is called the proximal point mapping associated with $f$.
\begin{proposition}[\citet{CuiL16}]\label{Prop1}
\rm Let $f:\cX\rightarrow (-\infty, +\infty]$ be a closed proper convex function, $ \varphi_{f}^{\rho}$ be the Moreau-Yosida regularization of $f$, and $P_{f}^{\rho}$ be the associated proximal point mapping. Then the following results hold.
\begin{enumerate}
  \item $ \varphi_{f}^{\rho}$ is continuously differentiable with gradient given by
  \begin{equation*}
    \nabla  \varphi_{f}^{\rho}(x)=\frac{1}{\rho}(x-P_{f}^{\rho}(x)).
  \end{equation*}
  Furthermore, $\nabla  \varphi_{f}^{\rho}$ is continuously Lipschitz continuous with modulus $1/\rho$.
  \item For any $x_1, x_2 \in\cX$, one has
  \begin{equation*}
    \langle P_{f}^{\rho}(x_1)-P_{f}^{\rho}(x_2),x_1-x_2\rangle\geq \|P_{f}^{\rho}(x_1)-P_{f}^{\rho}(x_2)\|^2,
  \end{equation*}
  which implies the mapping $P_{f}^{\rho}(\cdot)$ is globally continuous with modulus $1$ by the Cauchy-Schwarz inequality.
\end{enumerate}
\end{proposition}
The proof of Proposition \ref{Prop1} can be found in \citep[p.320]{HurL93}. Proposition \ref{Prop1} shows that the Moreau-Yosida regularization of any closed proper convex function is continuously differentiable, which is useful for the constrained optimization problem involving nonsmooth function.

\section{Derivation of \eqref{Estep}}
\label{sect.A}
\begin{proof}
Let $\bF=[\bff_1,\bff_2,\cdots,\bff_n]$ be treated as the missing data, $\bY=[\by_1,\by_2,\cdots,\by_n]$ be the observed one. Then the log-likelihood function of complete data is given by
\begin{align}\label{logl_h_f}
    L_c(\bLambda,\bSigma_{\be})=&-\frac{n}{2}\log(|\bSigma_{\be}|)-\frac{1}{2}\sum^{n}_{i=1}(\by_i-\bmu-\bLambda \bff_i)^T\bSigma_{\be}^{-1}(\by_i-\bmu-\bLambda \bff_i)-\frac{1}{2}\sum^{n}_{i=1}\bff_i^T\bff_i+\C
\end{align}
Since $\bar{\by}$ is the MLE of $\bmu$ and the conditional distribution of $\bff_i$ given $\by_i$ is
\begin{equation*}
\bff_i\sim \cN\left({\bGamma^{(k)}}^T(\by_i-\bar{\by}),\bOmega^{(k)}\right),
\end{equation*}
with $\bGamma^{(k)}={\bSigma_{\by }^{(k)}}^{-1}\bLambda^{(k)}$ and $\bOmega^{(k)}=I_r-{\bLambda^{(k)}}^T\bGamma^{(k)}$, the conditional expectation of $L_c(\bLambda,\bSigma_{\be})$ given $Y$ is
\begin{align}\label{Eq.ExpectLc}
  \E\left(L_c(\bLambda,\bSigma_{\be})|\bY\right) =& -\frac{n}{2}\log(|\bSigma_{\be}|)-\frac{n}{2} \tr(\bSigma_{\be}^{-1}\bS_{\by})-\frac{1}{2}\E\left(\sum^{n}_{i=1}\bff_i^T(\bLambda^T\bSigma_{\be}^{-1}\bLambda+I_r)\bff_i|\bY\right) \nonumber\\
   & + \E\left(\sum^{n}_{j=1}\bff_i^T\bLambda^T\bSigma_{\be}^{-1}(\by_i-\bar{\by})|\bY\right)+\C
\end{align}
With a little algebra, we get
\begin{eqnarray}\label{Eq.Expectdratic}
  &&\E\left(\sum^{n}_{j=1}\bff_i\bff_i^T|\bY\right) = n\left({\bGamma^{(k)}}^T\bS_{\by}\bGamma^{(k)}+\bOmega^{(k)}\right),\label{Eq.Expectdratic1}\\ &&\E\left(\sum^{n}_{j=1}(\by_i-\bar{\by})\bff_i^T|\bY\right) = n \bS_{\by}\bGamma^{(k)}.\label{Eq.Expectdratic2}
\end{eqnarray}
Substituting \eqref{Eq.Expectdratic1} and \eqref{Eq.Expectdratic2} into \eqref{Eq.ExpectLc} and adding the penalized term  give \eqref{Estep}.
\end{proof}

\section{Proof of Theorem \ref{Thm1}}
\label{sect.B}
\begin{proof}
For simplicity of presentation, we set $f$ be the generic density function. Let $f(\by|\bLambda,\bSigma_{\be})$ be the density function, then we use the following penalized maximum likelihood function
\begin{equation}\label{}
    L_{p}(\bLambda,\bSigma_{\be})=\log\left(f(\bY|\bLambda,\bSigma_{\be})\right)- \frac{n}{2}\mathrm{P}_{\lambda}(\bSigma_{\be}),
\end{equation}
where $\bY=[\by_1,\by_2,\cdots,\by_n]$. Given the current estimates $\bLambda^{(k)}$ and ${\bSigma_{\be}^{(k)}}$, the conditional expectation of the complete data log-likelihood function given $\bY$ is
\begin{align*}\label{}
    \E(L_c(\bLambda,\bSigma_{\be})|\bY)=& \int \log f(\bF,\bY|\bLambda,\bSigma_{\be})f\left(\bF|\bY,\bLambda^{(k)},\bSigma_{\be}^{(k)}\right)dF -\frac{n}{2}\mathrm{P}_{\lambda}(\bSigma_{\be})\nonumber\\
=&\int \log \left(f(\bF|\bY,\bLambda,\bSigma_{\be})f(\bY|\bLambda,\bSigma_{\be})\right) f\left(\bF|\bY,\bLambda^{(k)},\bSigma_{\be}^{(k)}\right)dF-\frac{n}{2}\mathrm{P}_{\lambda}(\bSigma_{\be})\nonumber\\
=&L_{p}(\bLambda,\bSigma_{\be})+\E\log\left(\frac{f(\bF|\bY,\bLambda,\bSigma_{\be})} {f\left(\bF|\bY,\bLambda^{(k)},\bSigma_{\be}^{(k)}\right)}\right)\\
    &+\E \log\left(f\left(\bF|\bY,\bLambda^{(k)},\bSigma_{\be}^{(k)}\right)\right).
\end{align*}
Since $-\log(\cdot)$ is convex, we have
\begin{align*}
&-\E\log\left(\frac{f(\bF|\bY,\bLambda,\bSigma_{\be})}{f\left(\bF|\bY,\bLambda^{(k)},\bSigma_{\be}^{(k)}\right)}\right) \geq-\log \E \left(\frac{f(\bF|\bY,\bLambda,\bSigma_{\be})}{f\left(\bF|\bY,\bLambda^{(k)},\bSigma_{\be}^{(k)}\right)}\right)=0
\end{align*}
 by Jensen's inequality. Hence, we get
\begin{align}\label{Eq.IncrP}
    \E(L_c(\bLambda,\bSigma_{\be})|\bY)&\leq L_{p}(\bLambda,\bSigma_{\be})+ \E \log\left(f\left(\bF|\bY,\bLambda^{(k)},\bSigma_{\be}^{(k)}\right)\right).
\end{align}
Given $\bSigma_{\be}^{(k)}$ and $\bLambda^{(k)}$, we update $\bLambda$ by minimizing $-\E(L_c(\bLambda, \bSigma_{\be}^{(k)})|\bY)$. Thus, we have
\begin{equation*}
    \E\left(L_c\left(\bLambda^{(k+1)}, \bSigma_{\be}^{(k)}\right)|\bY\right)\geq \E\left(L_c\left(\bLambda^{(k)}, \bSigma_{\be}^{(k)}\right)|\bY\right).
\end{equation*}
Then, $\bSigma_{\be}$ is updated by minimizing $-\E(L_c(\bLambda^{(k+1)}, \bSigma_{\be})|\bY)$. Since the MM procedure is nonincreasing and the APG algorithm is convergent, we get
\begin{equation*}
    \E\left(L_c\left(\bLambda^{(k+1)}, \bSigma_{\be}^{(k+1)}\right)|\bY\right)\geq \E\left(L_c\left(\bLambda^{(k+1)}, \bSigma_{\be}^{(k)}\right)|\bY\right).
\end{equation*}
According to \eqref{Eq.IncrP}, the following inequalities hold
\begin{align*}
  L_{p}\left(\bLambda^{(k+1)},\bSigma_{\be}^{(k+1)}\right)+ \E \log\left(f\left(\bF|\bY,\bLambda^{(k)},\bSigma_{\be}^{(k)}\right)\right)\geq & \E\left(L_c\left(\bLambda^{(k+1)}, \bSigma_{\be}^{(k+1)}\right)|\bY\right) \\
  \geq &\E\left(L_c\left(\bLambda^{(k)}, \bSigma_{\be}^{(k)}\right)|\bY\right)\\
  = &L_{p}\left(\bLambda^{(k)},\bSigma_{\be}^{(k)}\right)+ \E \log\left(f\left(\bF|\bY,\bLambda^{(k)},\bSigma_{\be}^{(k)}\right)\right).
\end{align*}
Just dividing the both sides of the above inequalities by $-n/2$, the nonincreasing property of the EMAPG algorithm follows easily.
\end{proof}

\end{document}